\documentclass[showpacs,preprintnumbers,amsmath,amssymb]{revtex4}
\usepackage{graphicx}
\usepackage{epic,eepic}
\usepackage{graphics}
\usepackage{epsfig}
\usepackage{subfigure}

\newcommand{\G}{{\cal G}}

\newcommand{\al}{\alpha}

\usepackage[latin1]{inputenc}
\usepackage{amsmath}
\usepackage{graphicx}
\usepackage{verbatim}

\usepackage{graphicx}
\usepackage{verbatim}

\newcommand{\be}{\begin{equation}}
\newcommand{\ee}{\end{equation}}
\newcommand{\bea}{\begin{eqnarray}}
\newcommand{\eea}{\end{eqnarray}}

\usepackage{amsmath}
\usepackage{amsfonts}
\usepackage{amssymb}
\usepackage{bm}
\usepackage{color}
\usepackage{graphicx}
\begin{document}

\title{Interaction Pressure Tensor for a class of Multicomponent Lattice Boltzmann models}

\author{M. Sbragaglia, D. Belardinelli} \affiliation{Department of Physics and INFN, University of Tor Vergata, \\ Via della Ricerca Scientifica 1, 00133 Rome, Italy} 

\begin{abstract}
We present a theory to obtain the pressure tensor for a class of non-ideal multicomponent lattice Boltzmann models, thus extending the theory presented by Shan (X. Shan, {\it Phys. Rev. E} {\bf 77}, 066702 (2008)) for single component fluids.  We obtain the correct form of the pressure tensor directly on the lattice and the resulting equilibrium properties are shown to agree very well with those measured from numerical simulations. Results are compared with those of alternative theories. 
\end{abstract}

\maketitle
\section{Introduction}

Modelling and simulation of multicomponent/multiphase fluid flows is extremely important and difficult, especially because of the problems in simulating complex diffusion processes, phase separation and interface dynamics \cite{Prosperetti,Brennen,Onuki}. The macroscopic behavior of the fluid system, including the dynamics of the coarse grained fields (density, velocity, temperature), is a direct consequence of the dynamics of the distribution functions at the atomistic level. Given the intrinsic microscopic roots of such phenomena, it is natural to design appropriate models describing mesoscopic physical interactions coupled to hydrodynamics, bypassing the need for interfacial treatment commonly required in other methods \cite{Prosperetti}. The lattice Boltzmann method, as one of such approaches, has been proven to be particularly suitable to the study of multicomponent and multiphase systems \cite{Benzi92,Chen98,Aidun10} where interfacial dynamics and phase transitions are present, since it can capture basic essential features of phase separation, even with simplified kinetic models. A significant progress has recently been made in this direction, as evidenced by many lattice Boltzmann models that have been developed on the basis of different points of view, including the Gunstensen model \cite{Gunstensen}, the Shan-Chen model \cite{SC,ShanDoolen}, the free-energy  model \cite{YEO}, and other kinetic models designed to simulate binary mixtures \cite{McCrackenAbraham05,Arcidiaconoetal07,Asinari}. 

Among these models, the potential model developed many years ago by Shan \& Chen \cite{SC,ShanDoolen} is widely used for its simplicity and efficiency in depicting interactions between different species and different phases \cite{Kupershtokh,Hyv,Sbragaglia06,Sbragaglia09,CHEM09,Sbragaglia09b,Sbragaglia07,Shan06b,Sbragagliaetal12,VarnikSaga,JansenHarting11}.  Controlling the equilibrium properties of such models is crucial for the application of the model itself in the description of capillary phenomena and phase separation \cite{Rowlinson,Onuki}.  In a recent paper, Shan \cite{Shan08} examined this issue and provided a general methodology for calculating the interaction pressure tensor for a non ideal gas system exhibiting multiple phases: this allows for a precise control of the equilibrium properties, such as bulk densities and surface tension at the non ideal interface. Using such analysis, one can better design and adjust the associated pseudopotentials to reproduce a free energy model based on a square gradient theory of equilibrium interfaces \cite{SbragagliaShan11}. To the best of our knowledge, such analysis is still missing for the case of a multicomponent fluid \cite{ShanDoolen}: this is the aim of the present paper.\\
The paper is organized as follows: in section \ref{sec:model} we present the basic ingredients of the lattice model and we detail the lattice theory for calculating the pressure tensor. In sections \ref{sec:equilibrium} and \ref{sec:check} we specialize to the case of a one dimensional interface and present numerical checks to validate the theory, also comparing our predictions with those of other existing theories. Conclusions follow in section \ref{sec:conclusions}.

\section{The non-ideal Multicomponent model}\label{sec:model}

Consider a mixture of $S$  components of monatomic gases. The molecular weight of the $\sigma$-th component is assumed to be unitary for simplicity. Let ${\bm x}$, $t$ and ${\bm \xi}$ be the position, time and velocity vectors, respectively. The single-particle distribution function of the $\sigma$-th component, $f^{(\sigma)}({\bm x},{\bm \xi},t)$, is defined such that $f^{(\sigma)}({\bm x},{\bm \xi},t) d{\bm x} \, d {\bm \xi}$ is the probability of finding a particle of the $\sigma$-th component in the element $d{\bm x} \, d {\bm \xi}$ of the phase space at time $t$.  By definition, the density $\rho$ and the fluid velocity ${\bm u}$  are moments of the distribution function, $\rho({\bm x},t)=\sum_{\sigma=1}^{S} \rho_{\sigma}({\bm x},t)=\sum_{\sigma=1}^{S}\int f^{(\sigma)}({\bm x},{\bm \xi},t) d {\bm \xi}$, ${\bm u}({\bm x},t)= \frac{1}{\rho}\sum_{\sigma=1}^{S}\int f^{(\sigma)}({\bm x},{\bm \xi},t) {\bm \xi} d {\bm \xi}$.  A multicomponent model on a discretized space-time lattice can then be derived: we first discretize the velocity space by projecting the dynamical equations into the Hilbert space spanned by the leading Hermite polynomials. A set of discrete velocities ${\bm \xi}_{\alpha}$ ($\alpha=1,\ldots,d$) is then chosen to coincide with the abscissas of a Gauss-Hermite quadrature in velocity space \cite{Shan06}.\\ 
Crucial for our present analysis is the characterization of the interaction part of the model. In presence of non-ideal effects, we may write the force experienced by the particles of the $\sigma$-th specie at ${\bm x}$, due to the particles of the other species at ${\bm y}$, in the following form:
$$
{\bm F}^{(\sigma)}({\bm x},{\bm y})= - \sum_{\sigma^{\prime} \neq \sigma} {\cal G}_{\sigma, \sigma^{\prime}}({\bm x},{\bm y})\psi_{\sigma}({\bm x})\psi_{\sigma^{\prime}}({\bm y})
({\bm y} - {\bm x})
$$
where $\psi_{\sigma}$ are the component-specific pseudopotentials, while ${\cal G}_{\sigma, \sigma^{\prime}}={\cal G}_{\sigma^{\prime}, \sigma}$ is a function which regulates the interactions between different pairs of components. It has to be noted that we have assumed that the interaction exists between different species only. When $\psi_{\sigma}=\rho_{\sigma}$, the model may be seen as a lattice transcription of a kinetic model with interactions arising only between particles of different species \cite{Basteaetal}. \\
When the sites interacting with the particles in ${\bm x}$ are limited to its $N$ neighbors (not necessarily the nearest ones), we may define a limited set of $N$ links as ${\bm c}_{\ell}$  ($\ell=0,\ldots,N-1$) which are not necessarily the same as those involved in the lattice Boltzmann streaming step (see section III). Requiring that the interaction is isotropic (i.e. all ${\bm x},{\bm y}$ such that $|{\bm y}-{\bm x}|=|{\bm c}_{\ell}|$ brings the same interaction strength), we then define ${\bm F}^{(\sigma)}_{\ell}({\bm x})= {\bm F}^{(\sigma)}({\bm x},{\bm x}+{\bm c}_{\ell})=-\sum_{\sigma^{\prime} \neq \sigma}{\cal G}_{\sigma, \sigma^{\prime}}(|{\bm c}_{\ell}|)\psi_{\sigma}({\bm x})\psi_{\sigma^{\prime}}({\bm x}+{\bm c}_{\ell}){\bm c}_{\ell}$ and sum over $\ell$ to obtain the total force experienced by particles of the $\sigma$-th specie in ${\bm x}$, due to the particles of the other species in neighbors locations:
\be\label{FORCE}
{\bm F}^{(\sigma)}({\bm x})=\sum_{\ell=0}^{N-1} {\bm F}^{(\sigma)}_{\ell}({\bm x})=-{\cal G}_{\sigma, \sigma^{\prime}}\psi_{\sigma}({\bm x})\sum_{\sigma^{\prime} \neq \sigma} \sum_{\ell=0}^{N-1} w_{\ell}\psi_{\sigma^{\prime}}({\bm x}+{\bm c}_\ell) {\bm c}_\ell
\ee
where the weights $w_{\ell}=w(|{\bm c}_\ell|^2)$ are defined by ${\cal G}_{\sigma, \sigma^{\prime}}(|{\bm c}_{\ell}|)={\cal G}_{\sigma, \sigma^{\prime}} w(|{\bm c}_\ell|^2)$. The momentum exchange introduced among lattice sites can be easily shown (for periodic boundary conditions) to sum up to zero, which ensures exact global momentum conservation \cite{SC}. For the sake of simplicity, we now make the assumption of two components ($\sigma=A,B$) and write $\G={\cal G}_{A,B}$ ($={\cal G}_{B,A}$). The forces on components $A$ and $B$ are, respectively,
\be\label{FORCEAB}
{\bm F}^{(A)}({\bm x})=\sum_{\ell=0}^{N-1} {\bm F}^{(A)}_{\ell}({\bm x})=-\G\psi_{A}({\bm x})\sum_{\ell=0}^{N-1} w_{\ell} \psi_{B}({\bm x}+{\bm c}_\ell) {\bm c}_\ell, \hspace{.2in} {\bm F}^{(B)}({\bm x})=\sum_{\ell=0}^{N-1} {\bm F}^{(B)}_{\ell}({\bm x})=-\G\psi_{B}({\bm x})\sum_{\ell=0}^{N-1} w_{\ell} \psi_{A}({\bm x}+{\bm c}_\ell) {\bm c}_\ell
\ee
with the link contributions ${\bm F}^{(A)}_{\ell}$, ${\bm F}^{(B)}_{\ell}$ given by:
$$
{\bm F}^{(A)}_{\ell}({\bm x})={\bm F}^{(A)}({\bm x},{\bm x}+{\bm c}_{\ell})=-\G  w_{\ell} \psi_{A}({\bm x})\psi_{B}({\bm x}+{\bm c}_{\ell}) {\bm c}_{\ell}, \hspace{.3in} {\bm F}^{(B)}_{\ell}={\bm F}^{(B)}({\bm x},{\bm x}+{\bm c}_{\ell})=-\G w_{\ell} \psi_{B}({\bm x}) \psi_{A}({\bm x}+{\bm c}_{\ell}) {\bm c}_{\ell}.
$$
\begin{figure}
\includegraphics[scale=1.0]{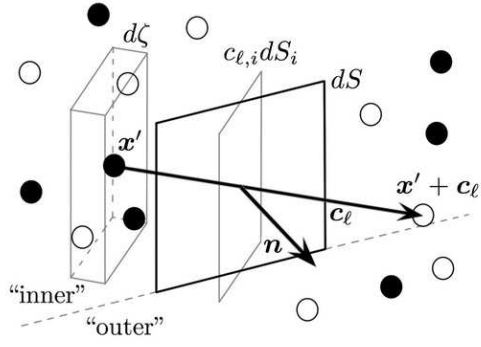}
\caption{A sketch showing some details for the computation of the pressure tensor. The infinitesimal surface $d {\bm S}$ divides the fluid in two parts: the ``outer'' part and the ``inner'' part. We then take the force $F_{j}^{(dS)}$ acting across $dS$ as the force acting on the particles ``inside $d{S}$'' due to the particles ``outside $d{S}$''. Interaction exists between different species only (filled and open symbols). The interaction pressure tensor ${P}_{i j}$ is then defined as ${F}^{(dS)}_{j}= -{P}_{i j} d {S}_{i}$.   \label{fig0}}
\end{figure}
Given the model for the lattice interactions (\ref{FORCE}), we next determine the associated pressure tensor \cite{Sbragaglia06,Shan08} responsible for mechanical balance at the interface between the two fluids. The idea is to connect the interaction forces ${\bm F}$ to the pressure tensor ${P}_{ij}$, by requiring that $-{P}_{ij} d{S}_{i}$ (double indexes are meant summed upon) is the $j$-th component of the total force acting across the surface $dS$ due to the interactions of particles on the opposite side. Here $d{\bm S}={\bm n}dS$, where ${\bm n}$ is the normal directed from the ``inner" to the ``outer" part of the fluid in a sense specified below. For the technical details, we will refer to a set of interaction links such that $|{\bm c}_{\ell}|^2 \le 1$ and $w_{\ell}=w(|{\bm c}_{\ell}|^2)=0$ for $|{\bm c}_{\ell}|^2>2$  \cite{Sbragaglia07,Shan06b}. This ensures isotropy for tensorial structures like $\sum_{\ell} w_{\ell} {c}_{\ell,i_1} {c}_{\ell,i_2} \ldots {c}_{\ell,i_k}$ up to the fourth order ($k=4$); more details can be found in Table \ref{T1}. 
\begin{table}
\begin{center}
\begin{tabular}{l l} 
\hline
\qquad \qquad Isotropy Weights (D2Q9) \\
\hline
$w_{\ell}  = 4/9  $&$\ell = 0$  \\
$w_{\ell}  = 1/9  $&$\ell = 1,4$  \\
$w_{\ell}  = 1/36 $&$\ell = 5,8$  \\
\hline 
\end{tabular} 
\begin{tabular}{l l} 
\hline
\qquad \qquad Isotropy Weights (D3Q19) \\
\hline
$w_{\ell}  = 1/3  $&$\ell = 0$  \\
$w_{\ell}  = 1/18  $&$\ell = 1,6$  \\
$w_{\ell}  = 1/36 $&$\ell = 7,18$  \\
\hline 
\end{tabular} 
\caption{Weights for the D2Q9 and D3Q19 models, ensuring isotropy up to the fourth order in the velocity tensors. More details can be found in  \cite{Shan06,Sbragaglia07}. \label{T1}}
\end{center}
\end{table}
Following Kirkwood \cite{Kirkwood50}, we then imagine a plane orthogonal to $d {\bm S}$ (that is, tangent to $dS$) in ${\bm x}$  and dividing the fluid in two parts: the ``outer'' part is just the portion into which the vector $d {\bm S}$ points, while the other part is the ``inner'' part. We then take, by convention, the force acting across $dS$ as the force acting on the particles ``inside $d{S}$'' due to the particles ``outside $d{S}$'', i.e. we consider those links  ${\bm c}_{\ell}$ for which ${c}_{\ell, i}{n}_{i}>0$ (see figure \ref{fig0}). The lattice forces are all between pairs of lattice sites. We then say that the force acts across $dS$ when the line connecting two interacting lattice sites intersects $dS$: if a particle of the $A$ ($B$) specie is located at ${\bm x}^{\prime}$ inside $dS$, and a particle of the $B$ ($A$) specie is located at ${\bm x}^{\prime}+{\bm c}_{\ell}$ outside $dS$, then the $j$-th component of the forces acting on $A$ ($B$) at ${\bm x}^{\prime}$ is 
\be\label{eq:f1}
{F}^{(A)}_{{\ell},{j}}({\bm x}^{\prime})=-\G w_{\ell} \psi_{A}({\bm x}^{\prime}) \psi_{B}({\bm x}^{\prime}+{\bm c}_{\ell}) {c}_{\ell ,j}, \hspace{.3in} {F}^{(B)}_{{\ell},{j}}({\bm x}^{\prime})=-\G w_{\ell} \psi_{B}({\bm x}^{\prime}) \psi_{A}({\bm x}^{\prime}+{\bm c}_{\ell}) {c}_{\ell ,j}.
\ee
These forces act across $dS$ only if ${\bm x}^{\prime}+\zeta {\bm c}_{\ell}$ goes across $dS$ for some $\zeta$ between $0$ and $1$. Fixing ${\bm c}_{\ell}$, the infinitesimal volume of the element around ${\bm x}^{\prime}$, over which ${\bm x}^{\prime}+ \zeta {\bm c}_{\ell}$ goes across $dS$ for  $\zeta$ in the interval [$\zeta$, $\zeta +d \zeta$] is ${c}_{\ell, i}d {S}_{i}  d \zeta$. The number of pairs of sites, one located in such a volume and another located in ${\bm x}^{\prime}+ {\bm c}_{\ell}$, is $\rho^{(2)}({\bm x}^{\prime},{\bm x}^{\prime}+ {\bm c}_{\ell}) \, {c}_{\ell, i}d {S}_{i}  \, d \zeta$, where $\rho^{(2)}$ indicates the number density of site pairs. The $j$-th component of the total force in the $\ell$-th direction acting across $dS$ is then given by the following integral in $d \zeta$
\[
-\G w_{\ell} {c}_{\ell, j} {c}_{\ell, i} d {S}_{i} \int_{0}^{1}[\psi_{A}({\bm x}-\zeta {\bm c}_{\ell}) \psi_{B}({\bm x}- \zeta {\bm c}_{\ell}+{\bm c}_{\ell}) + \psi_{B}({\bm x}-\zeta {\bm c}_{\ell}) \psi_{A}({\bm x}- \zeta {\bm c}_{\ell}+{\bm c}_{\ell})] \rho^{(2)}({\bm x}-\zeta {\bm c}_{\ell},{\bm x}- \zeta {\bm c}_{\ell}+{\bm c}_{\ell}) \, d \zeta.
\]
With the substitution $\gamma=1-\zeta$, using the symmetry of $\rho^{(2)}$ with respect to its two arguments, we see that the previous expression is equal to
\[
-\G w_{\ell} {c}_{\ell, j} {c}_{\ell, i} d {S}_{i} \int_{0}^{1}[\psi_{A}({\bm x}-{\bm c}_{\ell}+\gamma{\bm c}_{\ell}) \psi_{B}({\bm x}+\gamma{\bm c}_{\ell}) + \psi_{B}({\bm x}-{\bm c}_{\ell} + \gamma{\bm c}_{\ell}) \psi_{A}({\bm x}+\gamma{\bm c}_{\ell})]\rho^{(2)}({\bm x}+\gamma{\bm c}_{\ell},{\bm x}-{\bm c}_{\ell} +\gamma{\bm c}_{\ell}) \, d \gamma,
\]
which is the same as above but computed for $-{\bm c}_{\ell}$. Therefore, to obtain the $j$-th component of the total force acting on $dS$ in ${\bm x}$, we can sum over all $\ell$ (notice that we can also include terms for which $c_{\ell,i}n_{i}=0$) and divide by 2:
\be\label{eq:totalforce}
F_{j}^{(dS)}({\bm x})=-\frac{\G}{2} d {S}_{i} \sum_{\ell=0}^{N-1} w_{\ell}  {c}_{\ell, i} {c}_{\ell, j}\int_{0}^{1}[\psi_{A}({\bm x}-\zeta {\bm c}_{\ell}) \psi_{B}({\bm x}- \zeta {\bm c}_{\ell}+{\bm c}_{\ell}) + \psi_{B}({\bm x}-\zeta {\bm c}_{\ell}) \psi_{A}({\bm x}- \zeta {\bm c}_{\ell}+{\bm c}_{\ell})] \rho^{(2)}({\bm x}-\zeta {\bm c}_{\ell},{\bm x}- \zeta {\bm c}_{\ell}+{\bm c}_{\ell}) \, d \zeta.
\ee
It is then noted that for a system of ``pseudo'' particles localized on lattice sites, the pair density $\rho^{(2)}$ is a delta function peaked at the lattice sites and zero elsewhere. Assuming that ${\bm x}$ is one of the lattice sites, we see that the contribution to the integral arises from $\zeta=0$ and $\zeta=1$ only. We then use $\int_0^1 \delta(s)\, ds = \int_0^1 \delta(1-s)\, ds =1/2$ and the expression (\ref{eq:totalforce}) reduces to
\be\label{eq:totalforce2}
F_{j}^{(dS)}({\bm x}) = -\frac{\G}{2} d {S}_{i} \sum_{\ell=0}^{N-1} w_{\ell}  {c}_{\ell, i} {c}_{\ell, j}[\psi_{A}({\bm x}) \psi_{B}({\bm x}+{\bm c}_{\ell}) + \psi_{B}({\bm x}) \psi_{A}({\bm x}+{\bm c}_{\ell})].
\ee
We then use the definition of the interaction pressure tensor $P_{i j}$, given by ${F}^{(dS)}_{j}= -{P}_{i j} d {S}_{i}$, to get
\be\label{eq:PTlocal}
{P}_{i j}({\bm x})=\frac{\G}{2} \psi_A({\bm x}) \sum_{\ell=0}^{N-1} w_{\ell} \psi_B({\bm x}+{\bm c}_{\ell}) {c}_{\ell, i} {c}_{{\ell}, j}+\frac{\G}{2} \psi_B({\bm x}) \sum_{\ell=0}^{N-1} w_{\ell} \psi_A({\bm x}+{\bm c}_{\ell}) {c}_{{\ell}, i} {c}_{{\ell}, j}.
\ee
This, together with the ideal pressure contributions (intrinsic in the ideal gas dynamics) from the two components, gives the total pressure tensor
\be\label{eq:tot}
{P}^{(TOT)}_{i j}({\bm x})=c_s^2[\rho_A({\bm x})+\rho_B({\bm x})] I_{i,j} +\frac{\G}{2} \psi_A({\bm x}) \sum_{\ell=0}^{N-1} w_{\ell} \psi_B({\bm x}+{\bm c}_{\ell}) {c}_{\ell, i} {c}_{{\ell}, j}+\frac{\G}{2} \psi_B({\bm x}) \sum_{\ell=0}^{N-1} w_{\ell} \psi_A({\bm x}+{\bm c}_{\ell}) {c}_{{\ell}, i} {c}_{{\ell}, j}
\ee
with $c_s^2$ the square of the lattice sound speed and $I_{i,j}$ the identity tensor. With respect to the pressure tensor obtained for the single component fluid \cite{Shan08}, equation (\ref{eq:tot}) is more general and applies also to the case of a multicomponent fluid. We also note that expression (\ref{eq:tot}) has been derived for the case $|{\bm c}_{\ell}|^2 \le 1$ and $w_{\ell}=w(|{\bm c}_{\ell}|^2)=0$ for $|{\bm c}_{\ell}|^2>2$ (see table \ref{T1}). When the interaction links include higher order shells ($|{\bm c}_{\ell}|^2>2$), we refer the interested reader to \cite{Shan08} to obtain the generalizations of \eqref{eq:tot}.\\ 

\section{Equilibrium and Continuum Analysis}\label{sec:equilibrium}

To analyze the equilibrium properties of the model described in the previous section, we specialize the equation obtained for the pressure tensor \eqref{eq:tot}  to the case of a flat interface with spatial coordinate $x$ normal to the interface and unit lattice spacing. In presence of mechanical equilibrium ${P}^{(TOT)}_{xx}$ must be a constant. To obtain ${P}^{(TOT)}_{xx}$ we project the various contributions of equation (\ref{eq:tot}) along $x$
\be\label{PT1d}
P^{(TOT)}_{xx}(x)=c_s^2 \rho +{\cal G} \tilde{w} \psi_A(x) \frac{\psi_B(x+1)+\psi_B(x-1)}{2} + {\cal G} \tilde{w} \psi_B(x) \frac{\psi_A(x+1)+\psi_A(x-1)}{2}
\ee
where $\tilde{w}=1/6$ is a numerical constant coming from the average projection of the various $w_{\ell}$. A continuum limit for this pressure tensor can be found by substituting the Taylor expansion of $\psi_{A,B}(x \pm 1)$ into equation (\ref{PT1d}). Considering terms up to the second order derivatives we get
\be\label{1dPT}
P^{(TOT)}_{xx}(x) \approx \frac{1}{3} \rho+\frac{1}{3} {\cal G} \psi_A \psi_B+\frac{1}{6} {\cal G} \psi_A \frac{d^2 \psi_B}{dx^2} +\frac{1}{6} {\cal G} \psi_B \frac{d^2 \psi_A}{dx^2}
\ee
where we have used that $c_s^2=\sum_{\ell=0}^{N-1} w_{\ell} |{\bm c}_{\ell}|^2/D =1/3$ in (\ref{eq:tot}), with $D$ the space dimensionality. One may ask if equation (\ref{1dPT}) is consistent with the result one would get by solving the differential equation $-d P_{xx}/dx=F_x^{(A)}(x)+F_x^{(B)}(x)$, with the Taylor expansion (truncated at the third order derivatives) of the forcing $F_x^{(A,B)}(x)=-\tilde{w}{\cal G}\psi_{A,B}(x)(\psi_{B,A}(x+1)-\psi_{B,A}(x-1))$,  i.e.
$$
F_x^{(A)}(x)+F_x^{(B)}(x)=-\frac{1}{3}{\cal G}\psi_A(x)\frac{d\psi_B}{dx}-\frac{1}{18}{\cal G} \psi_A(x)\frac{d^3\psi_B}{dx^3}+A\leftrightarrow B. 
$$
The result, indicated with a $\prime$, is
 \be\label{1dPTcont}
P^{\prime (TOT)}_{xx}(x)=\frac{1}{3} \rho+\frac{1}{3} {\cal G} \psi_A \psi_B+\frac{1}{18} {\cal G} \psi_A \frac{d^2 \psi_B}{dx^2} +\frac{1}{18} {\cal G} \psi_B \frac{d^2 \psi_A}{dx^2} -\frac{1}{18} {\cal G} \frac{d\psi_A}{dx} \frac{d \psi_B}{dx}.
\ee
From its definition, it is clear that $P^{\prime (TOT)}_{ij}$ in (\ref{eq:tot}) is defined modulo a divergence-free tensor, say $A_{ij}$, with the property $\partial_j A_{ij}=0$. However, it seems that there is not a suitable choice of $A_{ij}$ which makes the two definitions (\ref{1dPTcont}) and (\ref{1dPT}) compatible: we see that we get the same bulk contribution $P_b=\frac{1}{3} \rho+\frac{1}{3} {\cal G} \psi_A \psi_B$ but different interfacial terms are present. The reason of such discrepancy is probably due to the cut-off introduced in the range of the lattice interactions. A similar calculation for single component fluids \cite{Shan08} reveals that the structure of the pressure tensor (\ref{1dPT}) is indeed affected by a mixed combination of first- and second-order derivatives of the pseudopotential, which appear as soon as the range of the interaction is increased. Nevertheless, for the practical use in a numerical simulation, relation (\ref{eq:tot}) represents an exact result and has an exact expression directly constructed on the lattice.

\section{Numerical Check}\label{sec:check}

\begin{figure}
\includegraphics[scale=0.7]{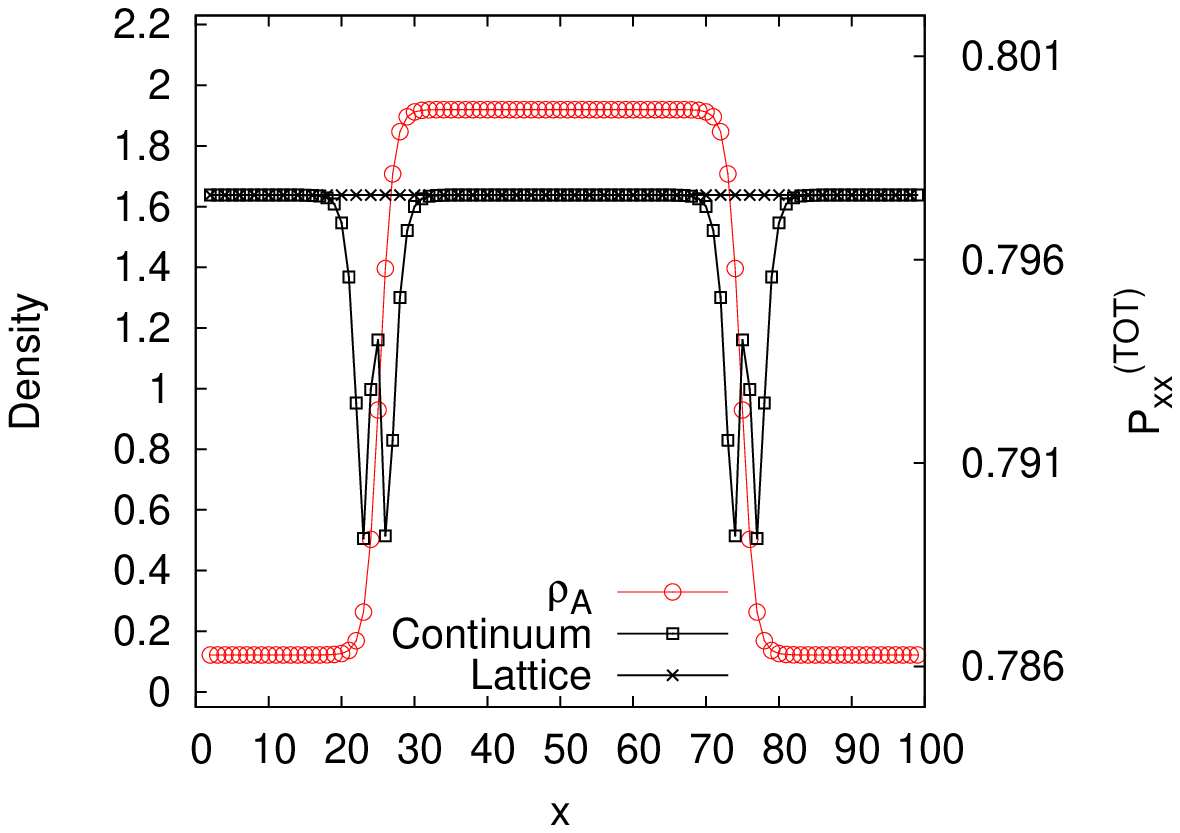}
\includegraphics[scale=0.7]{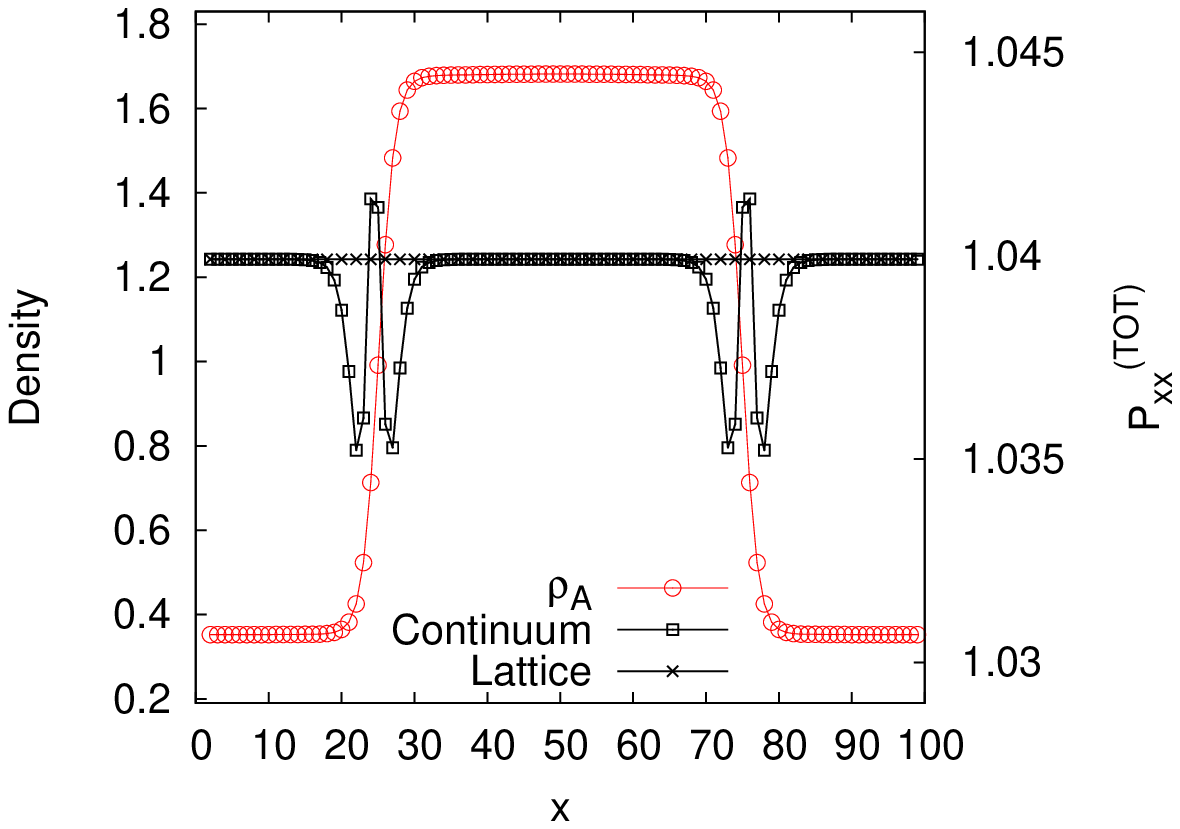}
\caption{Numerical simulations of a flat interface performed on a $L_x \times L_y = 100 \times 4$ domain where the hydrodynamical profiles are initialized so as to produce a one dimensional slab of width $L_x/2$. We plot the density of the $A$ component for a fixed $y=L_y/2$ at changing $x$ for the following choice of parameters: $\psi_{A,B}=\rho_{A,B}$, ${\cal G}=1.5$ leading to bulk densities $\rho_A=1.92$, $\rho_B=0.12$ in the $A$-rich region (left panel); $\psi_{A,B}=1-e^{-\rho_{A,B}}$, ${\cal G}=4.5$ leading to bulk densities $\rho_A=1.68$, $\rho_B=0.35$ in the $A$-rich region (right panel). Superimposed we report the lattice estimate of the pressure tensor ($\times$) as given in (\ref{PT1d}) and the continuum estimate ($\square$) provided by equation (\ref{1dPTcont}).  \label{fig:1}}
\end{figure}

\begin{figure}
\includegraphics[scale=0.7]{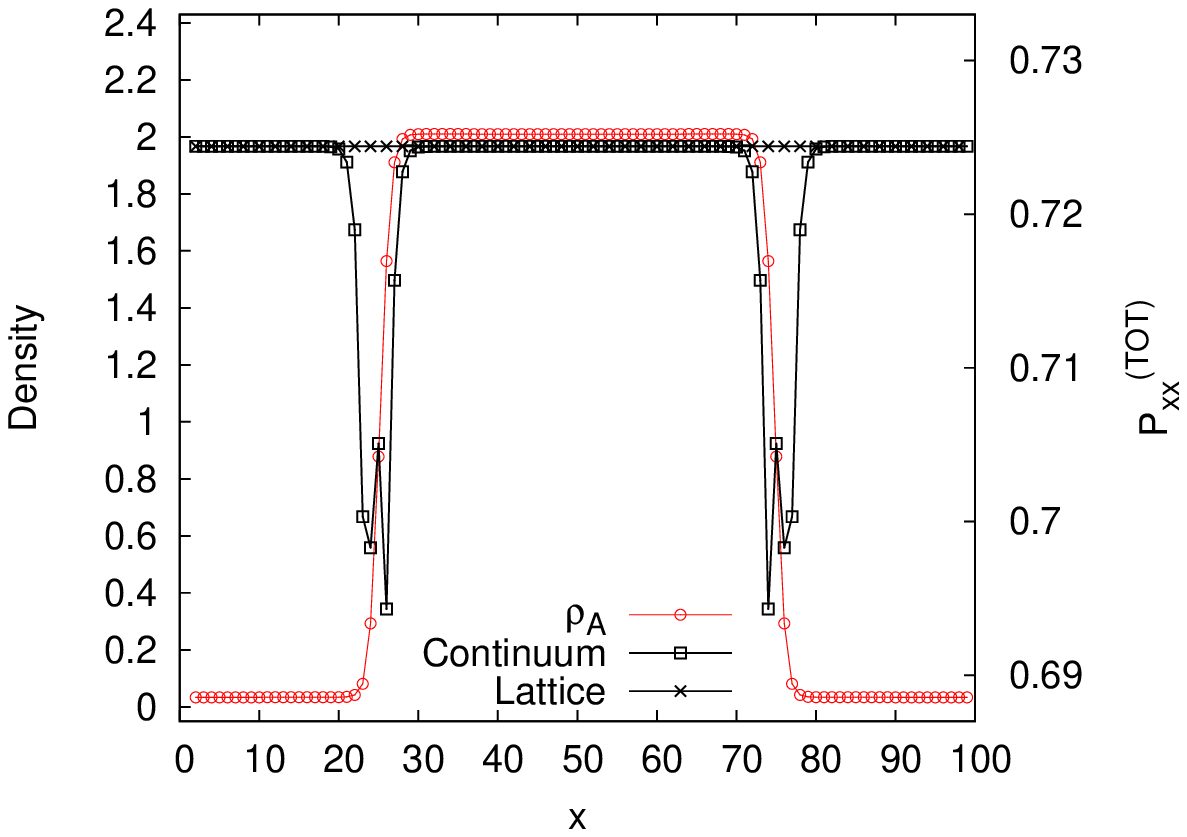}
\includegraphics[scale=0.7]{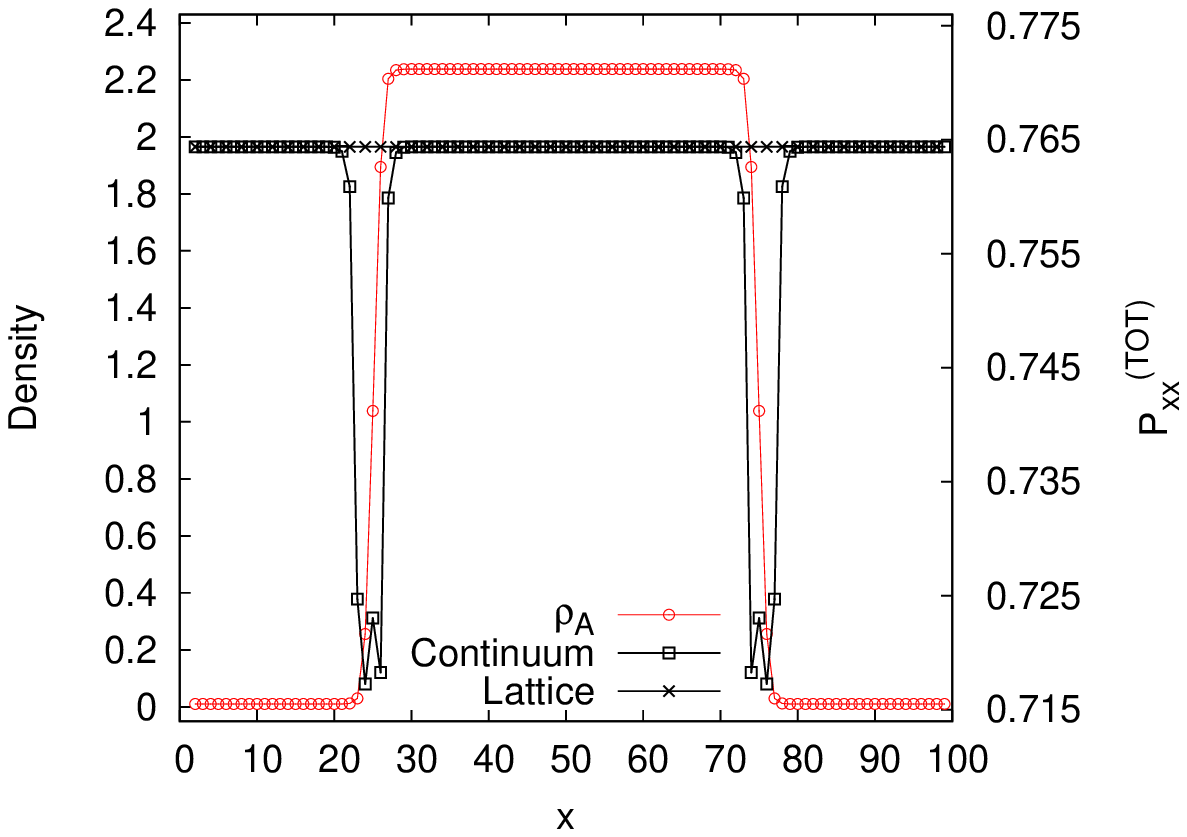}
\caption{Numerical simulations of a flat interface performed on a $L_x \times L_y = 100 \times 4$ domain where the hydrodynamical profiles are initialized so as to produce a one dimensional slab of width $L_x/2$. We plot the density of the $A$ component for a fixed $y=L_y/2$ at changing $x$ for the following choice of parameters: $\psi_{A,B}=\rho_{A,B}$, ${\cal G}=1.9$ leading to bulk densities $\rho_A=2.0$, $\rho_B=0.03$ in the $A$-rich region (left panel); $\psi_{A,B}=\rho_{A,B}$, ${\cal G}=2.0$ leading to bulk densities $\rho_A=2.24$, $\rho_B=0.01$ in the $A$-rich region (right panel). Superimposed we report the lattice estimate of the pressure tensor ($\times$) as given in (\ref{PT1d})  and the continuum estimate ($\square$) provided by equation (\ref{1dPTcont}).\label{fig:2}}
\end{figure}

For the numerical checks, we will use the D2Q9 lattice model with 9 velocities \cite{Succi} and a multiple-relaxation-time approximation (MRT) for the collisional operator \cite{Premnath}. A detailed description of the model can be found in other publications \cite{Guo,YuFan}. Here we briefly summarize its essential features. The evolution equation over a unit time lapse for such lattice Boltzmann scheme is
\be\label{DBEMRT}
f^{(\sigma)}_{\al}({\bm x}+{\bm \xi}_{\al},t+1)-f^{(\sigma)}_{\al}({\bm x},t)=- \Lambda^{\sigma}_{\alpha, \beta}  \left(f^{(\sigma)}_{\beta}-M_{\beta}(\rho_{\sigma},{\bm u}^{*}) \right) +  \left(I_{\alpha,\beta}-\frac{1}{2}\Lambda^{\sigma}_{\alpha, \beta}\right)  S_{\beta}({\bm u}^{*},{\bm F}^{(\sigma)})
\ee
with the internal force ${\bm F}^{(\sigma)}$ specified in equations (\ref{FORCEAB}). The Maxwellian equilibrium functions, truncated at the second order in the velocity vectors, are 
$$
M_{\alpha}(\rho,{\bm u}^*)=E_{\alpha} \rho \left[1+\frac{{\bm \xi}_{\al} \cdot {\bm u}^*}{c_s^2}+\frac{{\bm u}^*{\bm u}^*:({\bm \xi}_{\al} {\bm \xi}_{\al}-c_s^2 {\bm I} )}{2 c_s^4} \right]
$$
where the weights $E_{\alpha}$, similarly to the forcing weights, are chosen to enforce isotropy up to the fourth order in the velocity tensors (see Table \ref{T1}). The values $\Lambda^{\sigma}_{\alpha,\beta}$ are components of a $9 \times 9$ collision-scattering matrix whose eigenvalues are related to the transport coefficients of hydrodynamics. In particular, we choose a diagonal matrix with all the relaxation frequencies equal (which is equivalent to \cite{Guo}) to a unit value, i.e.  $\Lambda^\sigma_{\alpha,\beta}=I_{\alpha,\beta}$. The source term is chosen as
$$
S_{\alpha} ({\bm u}^{*},{\bm {F}}^{(\sigma)})=E_{\al}\left[\frac{1}{c_s^2}({\bm \xi}_{\al}-{\bm u}^*)+\frac{{\bm \xi}_{\al} \cdot {\bm u}^*}{c_s^4} {\bm \xi}_{\al} \right] \cdot {\bm {F}}^{(\sigma)}.
$$
In the above, ${\bm u}^*$ is the velocity of the whole fluid plus half of the total forcing contribution, i.e. the standard way to define the hydrodynamical velocity in the lattice Boltzmann scheme ${\bm u}^* ({\bm x},t) = {\bm u} ({\bm x},t)+\frac{{\bm F}}{2 \rho}({\bm x},t)$. \\
In figures \ref{fig:1} and \ref{fig:2}, we report the numerical simulations of a flat interface performed on a $L_x \times L_y = 100 \times 4$ domain, where the hydrodynamical profiles are initialized so as to produce a one dimensional slab of width $L_x/2$ extending from $L_x/4$ to $3 L_x/4$. After reaching the stationary state, we have first checked the validity of expression  (\ref{PT1d}) for two different pseudopotentials: in the first case (figure \ref{fig:1}, left panel) we set $\psi_{A,B}=\rho_{A,B}$; in a second case (figure \ref{fig:1}, right panel) we set $\psi_{A,B}=1-e^{-\rho_{A,B}}$ \cite{SC}. The coupling parameters are ${\cal G}=1.5$ and ${\cal G}=4.5$ for the first and second case, respectively. The $A$-rich phase has: $\rho_A=1.92$, $\rho_B=0.12$ (first case) and $\rho_A=1.68$, $\rho_B=0.35$ (second case), which are obtained by choosing a suitable average density in the initial condition. As we can see from figure \ref{fig:1}, the lattice pressure tensor (\ref{PT1d}) is constant throughout the interface, whereas the estimate given by the continuum pressure tensor (\ref{1dPTcont}) is not. The accuracy of the lattice pressure tensor is also appreciated  when the interface width is decreased and larger density ratios $|\rho_{A}-\rho_{B}|/(\rho_{A}+\rho_{B})$ in the bulk phases are achieved (see figure \ref{fig:2}). 


\section{Conclusions}\label{sec:conclusions}

We have used a statistical mechanics theory adapted to the discrete lattice Boltzmann dynamics of a multicomponent fluid to predict the associated pressure tensor. Such pressure tensor is found to be very accurate in characterizing mechanical equilibrium at non ideal interfaces.  An interesting topic for future research would be to control and predict the diffusive currents at the non ideal interface with expressions similar to equation (\ref{eq:PTlocal}). This would allow the complete control of the equilibrium (mechanical and chemical) directly on the lattice, instead of using continuum arguments as those presented in \cite{CHEM09}.\\

M. Sbragaglia kindly acknowledges funding from the European Research Council under the Europeans Community's Seventh Framework Programme (FP7/2007-2013) / ERC Grant Agreement no[279004].

\end{document}